\documentclass{ws-p8-50x6-00}

\def\Pom{{\bf I\!P}}

\def\lsim{\mathrel{\rlap{\lower4pt\hbox{\hskip1pt$\sim$}}
    \raise1pt\hbox{$<$}}}         %less than or approx. symbol

\def\gsim{\mathrel{\rlap{\lower4pt\hbox{\hskip1pt$\sim$}}
    \raise1pt\hbox{$>$}}}         %greater than or approx. symbol

\newcommand{\br}{{\bf r}}

\newcommand{\do}{\!\downarrow}

\begin{document}

\title{Connection between Diffraction and Small-$x$}

\author{N.N. Nikolaev}

\address{Institut f. Kernphysik, Forschungszentrum
J\"ulich\\ D-52425 J\"ulich, Germany\\
and\\
L.D. Landau Institute for Theoretical Physics\\ 142432 Chernogolovka, Russia
\\ 
E-mail: N.Nikolaev@fz-juelich.de}

%%%%%%%%%%%%%%%%%%%%%%%%%%%%%%%%%%%%%%%%%%%%%%%%%%%%%%%%%%%%%%
% You may repeat \author \address as often as necessary      %
%%%%%%%%%%%%%%%%%%%%%%%%%%%%%%%%%%%%%%%%%%%%%%%%%%%%%%%%%%%%%%

\maketitle

\abstracts{A brief review of the modern QCD theory of diffractive DIS
is given.}

\section{Why diffractive DIS is so fundamental}

Let us dream of e-Uranium DIS at THERA at $Q^2 \sim 10$ GeV$^2$ and $x \sim 10^{-5}$. 
Violent DIS is associated with complete destruction
of the target, deposition of a mere dozen MeV energy
breaks the uranium nucleus entirely, yet a rigorous
prediction from unitarity is that {\it diffractive} DIS $eU \to e'XU$ 
with the target nucleus emerging intact in the ground state will 
make $\approx 50\%$ of total DIS \cite{NZZnucleus}!
By the Glauber-Gribov theory, the abundance of diffraction
is closely related to nuclear shadowing (NS) in DIS. In 1974 Nikolaev 
and Zakharov reinterpreted NS in terms of the saturation of
nuclear parton densities \cite{NZ1975}. More recently, the NZ 
picture of saturation has been addressed to within QCD, see Iancu's talk 
at this Symposium \cite{IancuDatong}. If correct,
this QCD approach must inevitably lead to 50 \% diffractive DIS. 
To summarize, diffractive DIS
is a key to nuclear parton densities and QCD predictions 
for the initial state in ultrarelativistic nuclear collisions.
Because at HERA the rate of diffractive DIS is mere $10 \%$,
saturation effects are all but marginal, see \cite{NNZscan}
and talks \cite{GotsmanMaor} at this Symposium .

\section{Color dipole link between inclusive
and diffractive DIS}

\begin{figure}[!htb]
   \centering
   \epsfig{file=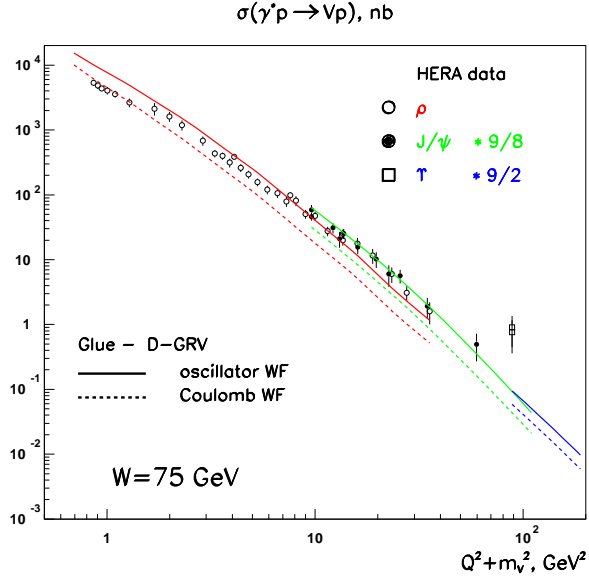,width=9cm}
\vspace{-0.5cm}
\caption{The test of the $(Q^2+m_V^2)$ scaling. The divergence of the 
solid and dashed curves indicates the sensitivity to the WF of the VM.
The experimental data are from HERA$^{~14}$.} 
\label{Figure1}
\end{figure}

\begin{figure}[!htb]
\vspace{-0.5cm}   
\centering
   \epsfig{file=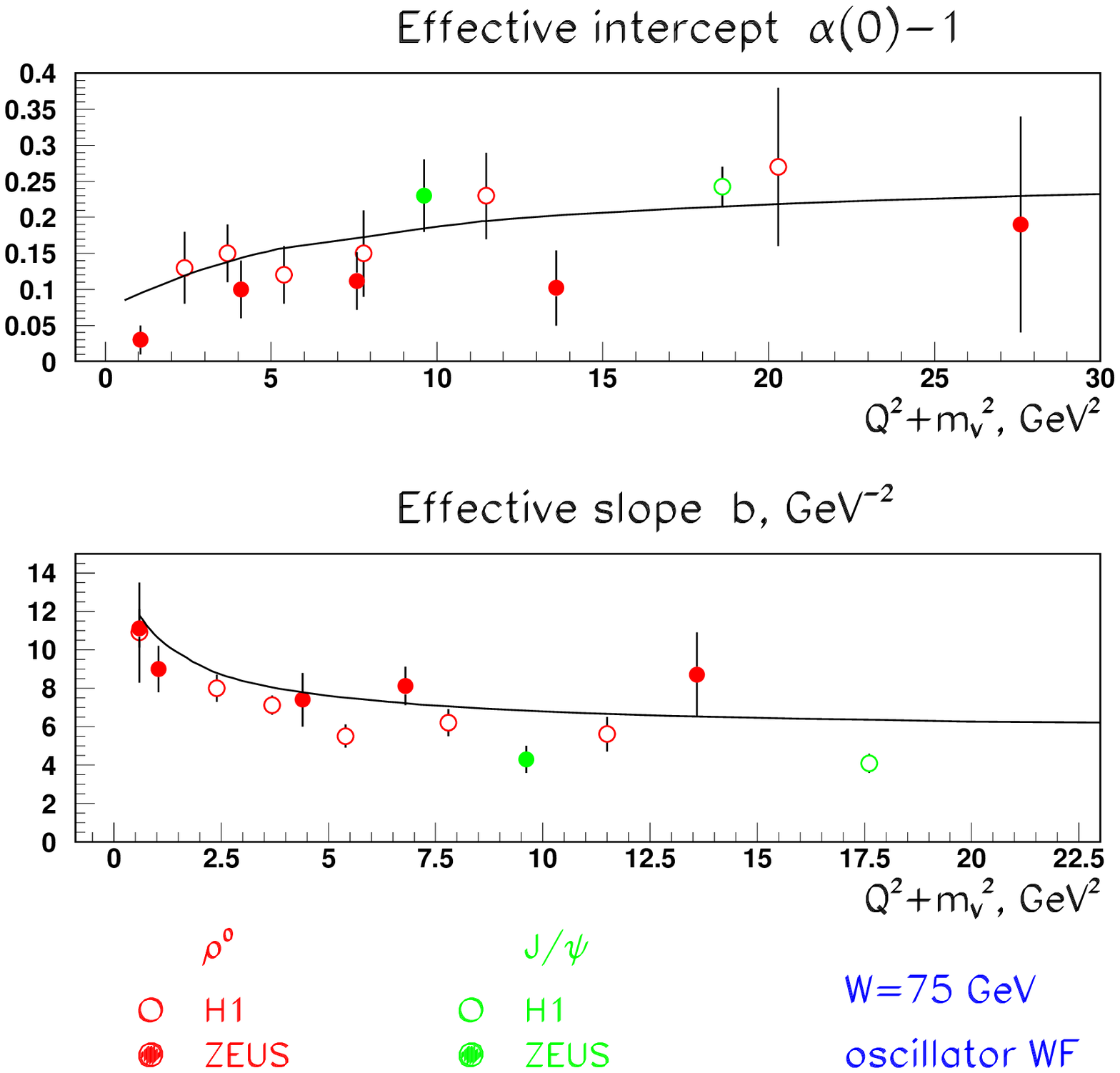,width=7cm}
\vspace{-0.5cm}
\caption{The $(Q^2+m_V^2)$ scaling of the effective intercept and 
diffraction slope$^{~17}$
}
\label{Figure2}
\end{figure}

SF's of DIS are related by optical theorem to the 
forward virtual Compton scattering (CS), which
in the color dipole (CD) factorization \cite{NZ91} takes the form
$
A_{CS}=\Psi^{*}_{f}\otimes A_{q\bar{q}}\otimes
\Psi_{in}
$
where $\Psi_{f,in}$ is the wave function (WF) of the $q\bar{q}$ 
Fock state of the photon and the  $q\bar{q}$-proton 
scattering kernel $A_{q\bar{q}}$ is proportional to the color dipole 
cross section, which for small dipoles is related to the gluon SF 
of the target, 
$
\sigma(x,\br)\approx {\pi^2 \over 3}r^2 \alpha_{S}({A\over r^2})
G(x,{A\over r^2})\, . 
%\label{eq:2.2}
$
where $A\approx 10$ by properties of the Bessel functions 
\cite{NZglue}. Taking for $\Psi^{*}_{f}$ the WF of the 
vector meson (VM) or the $X=q\bar{q}$ plane waves gives the diffraction 
excitation of VM \cite{NNZscan} or hadronic continuum, 
$\gamma^* p \to Xp'$. Generalization to excitation of the
$q\bar{q}g$ or higher Fock states of the photon 
is straightforward \cite{NZ92,NZ93}, with certain reservations
CD results can be reinterpreted in terms of the parton structure of
the pomeron with the solid result \cite{NZ92,NZ93,GNZ95} that gluons and 
charged partons carry about an equal fraction of the momentum of the
pomeron.
The formalism set in \cite{NZ92,NZ93} and especially in 
\cite{GNZ95,GNZcharm} is a basis of modern 
parameterizations of the diffractive structure function \cite{Bartels}.
 Unfortunately the use of the discredited
Ingelmann-Schlein-Regge factorization and DGLAP evolution
which is not warranted at large $\beta$ \cite{NZ93,GNZcharm},
in the still simplified form of this analysis
makes conclusions \cite{Bartels,DatongHERA} on the gluon content 
of the pomeron highly suspect.

%\begin{figure}[t]
%\figurebox{20pc}{15pc}{} % to have a box alone
%\epsfxsize=10pc % will enlarge or reduce the postscript figures based on the xsize
%\epsfbox{KolyaTriesteFig1.eps} % postscript image file name
%\caption{The test of the $(Q^2+m_V^2)$ scaling. The divergence of the 
%solid and dashed curves indicates the sensitivity to the WF of the VM.
%The experimental data are from the HERA experiments [...].}
%\end{figure}

\section{The $Q^2+m_V^2$ scaling}
While  DIS probes CD
$\sigma(x,\br)$ in a broad
range of ${1 \over AQ^2} \lsim r^2 \lsim 1 $ fm$^2$,
the diffractive VM 
production probes $\sigma(x,\br)$ at a 
scanning radius  \cite{NZ92,NNZscan} 
$
r\sim r_{S}= {6/ \sqrt{Q^2 + m_{V}^2}}\, ,
%\label{eq:2.2}
$
and the gluon SF of the target at the hard scale
$\overline{Q}^2 \approx$ (0.1-0.25)$* (Q^2 + m_{V}^2)$ 
and $x=0.5(Q^2+m_{V}^2)/(Q^2+W^2)$. After factoring out the 
charge-isospin factors, that entails the $(Q^2 + m_{V}^2)$ scaling 
of the VM production cross section\cite{NNZscan}, see fig.~1. The
same scaling holds also
for the effective intercept $\alpha_{\Pom}(0)-1$ of the energy
dependence of the production amplitude and
contribution to the diffraction slope $B$ from the $\gamma^* \to V$
transition vertex, which is $\propto r_{S}^2$ and exhibits the 
$(Q^2 + m_{V}^2)$ scaling \cite{NZZslope}, see fig.~2.
This $(Q^2 + m_{V}^2)$ scaling formulated in 1974, has recently become
a popular way of presenting the experimental data \cite{DatongVM}.
The theoretical calculations \cite{Igor} are based on the differential 
glue in the proton found in \cite{INdiffglue}

\section{Shrinkage of the diffraction cone in hard diffraction}

Gribov's shrinkage of the diffraction cone (DC)
$B = B_{0}+2\alpha'_{\Pom}\log W^2$, quantized in terms of
the slope  $\alpha'_{\Pom}$ of the pomeron trajectory, is the
salient feature of hadronic scattering, which derives from 
the Gribov-Feinberg-Chernavski diffusion in the
impact parameter space. 
At this Symposium, we heard of the ZEUS finding
\cite{DatongVM}
of the shrinkage of the DC in $\gamma p\to J/\Psi p$
with the result $\alpha'_{eff} =0.122\pm 0.033(stat)+0.018-0.032(syst)$
GeV$^{-2}$. Precisely such
a shrinkage has been predicted in 1995 by Nikolaev, Zakharov and
Zoller \cite{NZZslope} to persist within QCD even for
hard processes. 

First, in the usual approximation $\alpha_S=const$
the BFKL pomeron is the fixed cut in the complex-$j$ plane \cite{FKL}. 
Already in their first, 1975,  publication on QCD pomeron, Kuraev,
Lipatov and Fadin commented that incorporation of the asymptotic
freedom splits the cut into a sequence of  moving Regge poles \cite{FKL}, 
see Lipatov \cite{Lipatov} for more details.  
Within the CD approach, the Regge trajectories
of these poles where calculated in \cite{NZZslope,NZZpoles}. 
The CD cross section satisfies the CD BFKL equation \cite{NZZBFKL},
$\partial \sigma(x,r)/\partial\log{1\over x} = {\cal K}\otimes  \sigma(x,r)$,
which has the Regge solutions $\sigma_{n}(x,r) =\sigma_{n}(r)x^{-\Delta_{n}}$.
The CD kernel ${\cal K}$ is related to the flux of Weizs\"acker-Williams gluons
around the $q\bar{q}$ dipole. 
The NZZ strategy was to evaluate $\alpha_{n}'$ from the energy dependence
of $\lambda(x,r) = B\sigma(x,r)$, which satisfies the
inhomogeneous equation $\partial \lambda(x,r)/\partial\log{1\over x} - 
{\cal K}\otimes  
\lambda(x,r) = {\cal L}\otimes r^2 \sigma(x,r)$, and has solutions
$\lambda_{n}(x,r) =\sigma_{n}(x,r)\cdot \alpha'_{n} \log{1\over x}.$
Because $B= {1\over 2}
\langle b^2\rangle $
and the impact parameter $b$ receives a contribution from the gluon-$q\bar{q}$
separation $\rho$, the inhomogemeous term 
${\cal L}\otimes r^2 \sigma(x,r)$ is driven by precisley 
the impact parameter diffusion of WW gluons. Because 
$\alpha'_{n}$ is driven by the inhomogeneous term, there is 
a manifest relationship between $\alpha'_n$ and the 
Gribov-Feinberg-Chernavski diffusion in the
impact parameter space. Evidently, the dimensionfull quantity $\alpha'_{n}$
depends on the infrared regularization of QCD, within the specific
regularization \cite{NZZBFKL} which has lead to an extremely successful
description of the proton structure function (see \cite{NSZpion} and references therein),
for the rightmost hard BFKL pole we found $\alpha'_{\Pom}\approx 0.07$ GeV$^-2$.
The contribution from subleading BFKL poles was found to be still 
substantial at subasymptotic energy of HERA with the result
$\alpha_{eff}'\approx 0.15-0.17$ GeV$^{-2}$ \cite{NZZslope}, this
prediction from 1995 agrees perfectly with the ZEUS finding.

\section{Conclusions}

The QCD theory of diffractive DIS is gradually coming of age.
The fundamental $(Q^2+m_V^2)$ scaling predicted in 1994 has
finally been recognized by experimentalists. 
The shrinkage of the diffraction cone for hard photoproduction
$\gamma p \to J/\Psi p$ discovered by the ZEUS collaboration
is the single most important result. It shows that the BFKL
pomeron is a (series of) moving pole(s) in the complex-$j$
plane. The slope of the pomeron trajectory and the rate of
shrinkage of the diffraction cone for hard photo- and
electroproduction predicted in 1995 has been confirmed
experimentally. 

I'm grateful to the Symposium organizers, especially 
Profs. Liu Lianshou and  J.~Crittenden, 
for the invitation to ISMD31. This work was partly supported by
the INTAS grants 97-30494 and 00-00366.

\end{document}